%% file: Levinetal2013ver2.tex
\documentclass[useAMS,usenatbib]{mn2e}
\usepackage{times, graphicx, setspace, subfigure, latexsym, amssymb, amsmath, fancyhdr}
\usepackage{natbib}
\usepackage{multirow}
\usepackage{aastex_hack-test}
\title[HTRU VIII: The Galactic millisecond pulsar population]{The High Time Resolution Universe Pulsar Survey VIII: The Galactic millisecond pulsar population}
\author[L. Levin et al.]{L. Levin$^{1, 2, 3, 4}$\thanks{E-mail:lina.s.levin@gmail.com}, 
M. Bailes$^{1,3}$, B. R. Barsdell$^{1,3}$, S. D. Bates$^4$, N. D. R. Bhat$^{3,5}$, M. Burgay$^6$,
\newauthor
 S. Burke-Spolaor$^7$, D. J. Champion$^8$, P. Coster$^{1,2}$, N. D'Amico$^{6,9}$, A. Jameson$^{1,3}$, 
\newauthor
S. Johnston$^2$,  M. J. Keith$^2$, M. Kramer$^{8,10}$, S. Milia$^{6,9}$, C. Ng$^8$, A. Possenti$^6$, B. Stappers$^{10}$, 
\newauthor
D. Thornton$^{10}$ and W. van Straten$^{1,3}$\\
$^{1}$Swinburne University of Technology, Centre for Astrophysics and Supercomputing Mail H30, PO Box 218, VIC 3122, Australia\\
$^{2}$Australia Telescope National Facility, CSIRO Astronomy \& Space Science, P.O. Box 76, Epping, NSW 1710, Australia\\
$^3$ARC Centre of Excellence for All-Sky Astronomy (CAASTRO)\\
$^4$Department of Physics, West Virginia University, 210E Hodges Hall, Morgantown, WV 26506, USA\\
$^5$International Centre for Radio Astronomy Research, Curtin University, Bentley, WA 6102, Australia\\
$^6$INAF-Osservatorio Astronomico di Cagliari, localit\'{a} Poggio dei Pini, strada 54, I-09012 Capoterra, Italy\\
$^7$NASA Jet Propulsion Laboratory, M/S 138-307, Pasadena, CA 91106, USA\\
$^8$Max Planck Institut f\"{u}r Radioastronomie, Auf dem H\"{u}gel 69, 53121 Bonn, Germany\\
$^{9}$Dipartimento di Fisica, Universit\'{a} degli Studi di Cagliari, Cittadella Universitaria, 09042 Monserrato (CA), Italy\\
$^{10}$University of Manchester, Jodrell Bank Centre for Astrophysics, Alan Turing Building, Manchester M13 9PL, UK}

\begin{document}

\date{Accepted ... Received ...; in original form ...}

\maketitle

\begin{abstract}
We have used millisecond pulsars (MSPs) from the southern High Time Resolution Universe (HTRU) intermediate latitude survey area to simulate the distribution and total population of MSPs in the Galaxy. Our model makes use of the scale factor method, which estimates the ratio of the total number of MSPs in the Galaxy to the known sample.
Using our best fit value for the z-height, $z$\,=\,500\,pc, we find an underlying population of MSPs of 
$8.3 (\pm 4.2)\times10^4$ sources down to a limiting luminosity of $L_{\rm min}$\,=\,0.1\,mJy\,kpc$^{2}$ and a luminosity distribution with a steep slope of $d\log N/d\log L$ = --1.45$\pm0.14$. However, at the low end of the luminosity distribution, the uncertainties introduced by small number statistics are large. By omitting very low luminosity pulsars, we find a Galactic population above $L_{\rm min}$\,=\,0.2\,mJy\,kpc$^{2}$ of only $3.0 (\pm 0.7)\times 10^4$ MSPs.  
We have also simulated pulsars with periods shorter than any known MSP, and estimate the maximum number of sub-MSPs in the Galaxy to be $7.8 (\pm 5.0)\times 10^4$ pulsars at $L$\,=\,0.1\,mJy\,kpc$^{2}$. 
In addition, we estimate that the high and low latitude parts of the southern HTRU survey will detect 68 and 42 MSPs respectively, including 78 new discoveries. 
Pulsar luminosity, and hence flux density, is an important input parameter in the model. 
Some of the published flux densities for the pulsars in our sample do not agree with the observed flux densities from our data set, and we have instead calculated average luminosities from archival data from the Parkes Telescope. We found many luminosities to be very different than their catalogue values, leading to very different population estimates. Large variations in flux density highlight the importance of including scintillation effects in MSP population studies.
\end{abstract}


\section{Introduction}
To date pulsar astronomers have discovered over 2000 pulsars in the Galaxy and globular clusters, including over 170 millisecond pulsars (MSPs), 
and the known population continues to grow with the results from each new successful pulsar survey. However, the distribution of the observed sample implies that we only know of a small fraction of the total Galactic pulsar population. With the help of simulations we can use the properties of the known sample to estimate the underlying Galactic population.  

\subsection{Previous Studies}
Studies of the MSP population in the Galaxy were for a long time severely affected by small-number statistics. To some degree that is still true today, however with the new large scale surveys increasing the total numbers of known MSPs, we are quickly getting closer to being able to perform full population synthesis studies of MSPs with good accuracy.

The first attempt to estimate the number of MSPs in the Galaxy was made by \cite{kul88}, who examined the hypothesis that the birthrates of low-mass binary pulsars (LMBP) and low-mass X-ray binaries (LMXB) should be equal. This hypothesis is derived from the model of formation of MSPs through recycling \citep{alp82,fab83,van86}, where MSPs are ordinary pulsars spun-up by the accretion of matter from a companion. During the spin-up phase, the binary system is visible as an LMXB. \cite{kul88} argued that the birthrates of these two groups are in fact not equal, but that the birthrate of short orbital period LMBPs is exceeding the birthrate of LMXBs by a factor of $\sim$10. As part of their analysis they used 2 MSPs and estimated a total of $>$100,000 MSPs in the Galaxy. Their analysis was however dominated by one of these MSPs: PSR\,B1855+09. 
After the results from subsequent surveys by \cite{nar90} the number of pulsars inferred in the analysis was revised and limited to a smaller number, but the conclusion that the birthrate of MSPs is more than 10 times that of LMXBs was still present. 
This issue is sometimes referred to as the birthrate problem, and the question whether all MSPs are produced in LMXB systems is still discussed today \citep[e.g.][]{hur10}. 

In the early 1990s \cite{joh91} used five MSPs, including PSR\,B1855+09, and non-detections from two high-frequency surveys of the Galactic plane to estimate the disc population of MSPs. By using a revised distance scale, which increased the distance to PSR\,B1855+09, they attained a higher luminosity of the pulsar and hence derived a lower number of PSR\,B1855+09-like pulsars in the Galaxy. 
They estimated a total of $\sim 2\times10^5$ MSPs down to a limiting radio luminosity of $L_{\rm min, 1500MHz} = 0.3$\,mJy\,kpc$^2$ in the Galaxy. However, this result had a large uncertainty due to the small number of MSPs known at the time, and in addition, the result was dominated by the MSP B1257+12. 

One important result from the study by \cite{joh91} was their prediction that, in contrast to slow pulsars, the spatial distribution of MSPs would not be concentrated on the Galactic plane but evenly distributed over the sky. This understanding together with the first discoveries of MSPs at high galactic latitudes \citep{wol91} encouraged large-scale (nearly) all-sky surveys for MSPs to be undertaken at various radio telescopes \citep{tho93,man96,say97}. 

One of these surveys, the Parkes Southern Pulsar Survey \citep{man96}, discovered 17 MSPs and almost doubled the known sample resulting in 35 known MSPs at the time. Using these 17 discoveries together with the parameters for another 4 previously known MSPs, \cite{lyn98} estimated the local surface density of MSPs to be 1110$\pm$600 sources within a 1.5-kpc cylindrical radius of the Sun. This number corresponds to a local surface density of MSPs of 157$\pm$85\,kpc$^{-2}$ for $L_{\rm 436\,MHz} > 0.3$\,mJy\,kpc$^{2}$. 

A more recent study of the MSP population was reported by \cite{sto07}, who used 56 radio-loud MSPs from 10 different pulsar surveys to predict the number of radio-loud and radio-quiet MSPs that are detectable as $\gamma$-ray pulsars. By assuming that ordinary and millisecond pulsars all can be described with a common radio luminosity model, their extensive study results in a Galactic birth rate of MSPs of 4\,-\,5$\times 10^{-4}$ per century, which corresponds to a total Galactic population of 4.8\,-\,6.0$\times 10^4$ MSPs.

\subsection{Selection Effects in Pulsar Surveys}
The known pulsar population for both ordinary pulsars and MSPs is strongly biased towards bright sources at small distances from the Earth. 
Pulsar astronomers define the intrinsic luminosity of a source at the observing frequency $\nu$ as $L_{\nu} \equiv S_{\nu} d^2$, where $S_{\nu}$ is the mean flux density at $\nu$ and $d$ is the distance to the pulsar. The distance is usually derived from the dispersion measure (DM). This inverse square law results in a known sample dominated by nearby pulsars and those with high luminosity. 
Throughout this paper, $L$ refers to the luminosity at an observing frequency of 1400\,MHz. 

Other selection effects of pulsar surveys that affect the observed population are interstellar dispersion and scattering of pulses. 
To estimate the underlying population of pulsars it is important to keep these biases in mind and take them into account in any population modelling. 

The detection threshold of the apparent flux density for pulsar surveys is calculated with the radiometer equation by:
\begin{equation}
	S_{\rm min} = \frac{S/N_{\rm min}(T_{\rm rec}+T_{\rm sky})}{G \eta \sqrt{n_{\rm pol} t_{\rm int} \Delta \nu}}\sqrt{\frac{W}{1-W}} \quad {\rm mJy}
	\label{eq:smin}
\end{equation}
where $S/N_{\rm min}$ is the threshold signal-to-noise ratio, $T_{\rm rec}$ and $T_{\rm sky}$ are the receiver and sky noise temperatures (measured in K), $G$ is the telescope antenna gain (K/Jy), $\eta$ is a survey-dependent constant ($\le 1$) which accounts for losses in sensitivity due to e.g. sampling and digitization noise, $n_{\rm pol}$ is the number of polarizations recorded, $t_{\rm int}$ is the integration time (seconds), $\Delta \nu$ is the observing bandwidth (MHz) and $W$ is the observed pulse width given in parts of the pulse period \citep{dew85}.
From eq \ref{eq:smin} we can see that the minimum detectable flux density increases as $W$ increases, hence it is harder to detect pulsars with broad pulse profiles than those with narrower ones. 

Dispersion smearing and multi-path scattering by the free electrons in the interstellar medium cause the detected pulse width to be broader than the intrinsic pulse value. Since the density of free electrons is higher closer to the Galactic plane, these effects are more severe for distant pulsars in the inner Galaxy. These effects are also highly dependent on observing frequency and are less severe at higher observing frequencies ($\ge$ 1.4 GHz) than at frequencies around 400 MHz. The sky temperature scales with observing frequency \citep[approximately as $\nu^{-2.8}$;][]{law87} and makes higher observing frequencies attractive. On the other hand, pulsar flux densities in general possess a steep negative spectral index which causes the flux density to be roughly an order of magnitude lower at 1.4\,GHz compared to at 400\,MHz. This issue can be partly compensated for by increasing the receiver bandwidth at higher radio frequencies.

\section{Millisecond Pulsar Data Set}
\label{sec:mspdataset}
As input in the pulsar population model we have used recycled pulsars in the region of the sky covered by the intermediate latitude part of the southern High Time Resolution Universe  survey \citep[HTRU;][]{kei10}. The HTRU survey is an all-sky search for pulsars and fast transients currently underway at the Parkes 64-m Radio Telescope in Australia and at the Effelsberg 100-m Radio Telescope in Germany. The southern survey uses the 20-cm multibeam receiver \citep{sta96} at Parkes. The 13 beams in the multibeam receiver are mounted in a solid pattern consisting of a centre feed surrounded by two hexagonal rings of feeds with different sensitivity. The beam ellipticity and gain degradation of each feed are listed in Table \ref{tab:MBspecs}.
The data are sampled using 2 bits every 64\,$\mu$s, with an effective bandwidth of 341\,MHz, centred around 1.35\,GHz and divided into 874 frequency channels. For effective use of observing time, the survey region is divided up in three sub-surveys with different integration times and sky coverage. The data set used in this paper is based on observations from the intermediate latitude part of the southern survey (medlat), which covers a region of the sky limited by $-120^{\circ} < l < 30^{\circ}$, $|b| < 15^{\circ}$ with 540\,s integrations. 
The other two sub-surveys cover the low latitudes (deep) and the high latitudes (hilat) of the southern sky respectively. The deep survey is bounded by $|b| < 3.5^{\circ}$ and $-80^{\circ} < l < 30^{\circ}$ and has an integration time of 4300\,s. The hilat survey covers all the sky south of a declination of $+10^{\circ}$, not included in the medlat survey, with 270\,s integrations. 
More detail on the HTRU survey can be found in \cite{kei10}.

\begin{table}
  \caption[Multibeam receiver specifications]{Multibeam receiver specifications, showing the centre feed, the inner ring and outer ring of feeds. Values are taken from \cite{man01}.}
  \begin{center}
  \begin{tabular}{l l l l}
	\hline 
	\hline
	Beam & Centre & Inner ring & Outer ring\\
	\hline
	Telescope gain (K Jy$^{-1}$) & 0.735 & 0.690 & 0.581\\
	Half-power beamwidth (arcmin) & 14.0 & 14.1 & 14.5\\
	Beam ellipticity & 0.0 & 0.03 & 0.06\\
	Coma lobe (dB) & none & --17 & --14\\
	\hline
	\hline
  \end{tabular}
  \label{tab:MBspecs}
  \end{center}
\end{table}

For this analysis we have used 50 previously known and newly discovered recycled pulsars in the HTRU medlat survey area not associated with globular clusters. We define recycled pulsars as having a period $P < 0.070$\,s and a spin down rate $\dot{P} < 10^{-17}$s/s. All pulsars in our sample are listed in Table \ref{tab:mergetab} together with some of their properties and a note of in which 20-cm pulsar surveys they have been detected. 
Three of the newly discovered pulsars from the HTRU medlat survey were confirmed only shortly before this work started. Timing observations of these pulsars are carried out at Jodrell Bank Observatory and due to a lack of data from the Parkes Telescope at the time, these pulsars are not included in this analysis. 

To make sure the numbers of detected pulsars from our simulation in each survey was correct, archival data from the Parkes Multibeam survey (PM) and the Swinburne Intermediate Latitude survey (SIL) were searched for the sample pulsars. This was done by finding the observation in each survey closest on the sky to each of the pulsars. The data files were then de-dispersed at the DM of the known pulsar and searched for periodicities close to the known pulsar period by performing an FFT with the program {\it seek}\footnote{Part of the {\sc sigproc} package: http://sigproc.sourceforge.net/}. 
If a periodicity within $\pm$\,1\,$\mu$s of the known pulsar period was found, the data were folded using this periodicity and the known pulsar DM. 
The resulting folded archives were then inspected by eye to determine if the pulsar was detected in the data or not. The analysis resulted in an additional five MSP detections in the PM survey and three MSP detections in the SIL survey, compared to published values. All the new detections were of pulsars originally discovered in the HTRU medlat survey. 

Two of the previously known pulsars (J1157--5112 and J1454--5846) were not detected in the HTRU data. The positions for these pulsars were both at the edge of the covering beam in the survey. For PSR\,J1454--5846, the position offset reduces the average S/N enough for the source to be below the detection threshold of the medlat survey. That is not the case for PSR\,J1157--5112, which at its average flux density should be detectable even after corrections for the position offset is taken into account. However, this source has a low DM of $\sim 40$ and is highly affected by scintillation, which could explain the HTRU non-detection. 

The total numbers of pulsars from our sample detected in each survey are 19 pulsars in the PM survey, 15 in the SIL survey and 48 in the HTRU medlat survey.

\input{tables/inputvalues_ver2}

\begin{figure*}
	\begin{center}
		\includegraphics[height=0.9\textheight]{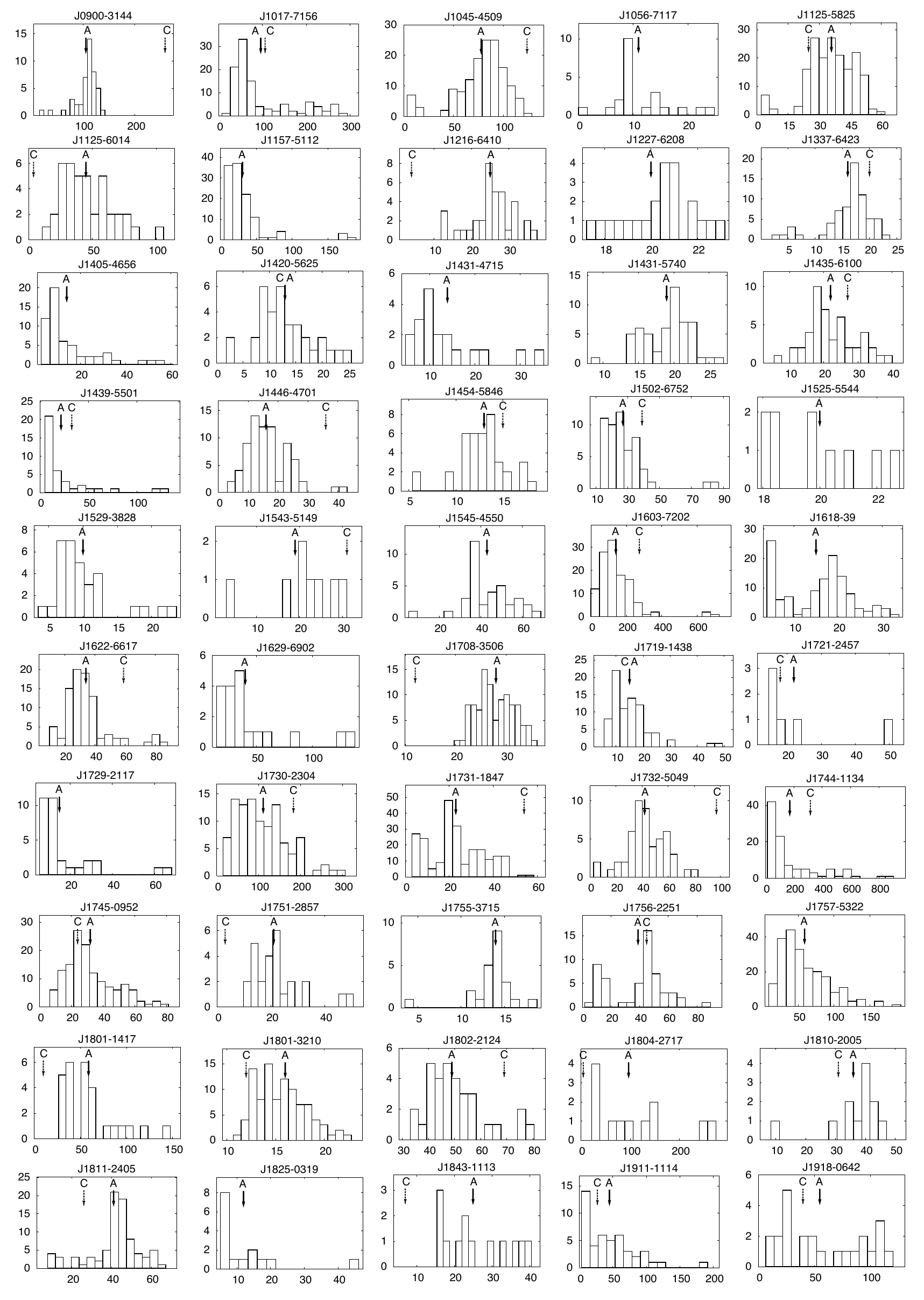}
	\caption[S/N histograms]{Histograms of S/N values for the sample MSPs from archival Parkes observations. The arrows marked A and C point to the values for the average S/N from the data set and the catalogue S/N derived from the published 20-cm flux density respectively. Where no C arrow is given, no published flux density value exists.}
	\label{fig:snrhist}
	\end{center}
\end{figure*}

\subsection{Pulsar Flux Densities}

Through the course of this analysis it became clear that the published flux densities for some of the pulsars as stated in the ATNF pulsar catalogue\footnote{http://www.atnf.csiro.au/research/pulsar/psrcat/} \citep{man05} do not agree with the values obtained in the observations in the HTRU medlat survey. In particular the values from four pulsars stood out as being a lot higher than the published flux density. The most obvious reason for this discrepancy would be if the pulsar signal was amplified by scintillation in our observation, and would then appear to be brighter than its average state. Since the luminosity of a pulsar is proportional to its flux density, it is very important to get the flux density values right before using the luminosity to derive the underlying pulsar sample. 

In an effort to better understand these differences between our data and the published flux density values, we have analysed a large number of archival observations collected with Parkes of the MSPs in our sample\footnote{Most of these observations are available via the CSIRO Data Access Portal: https://data.csiro.au/dap/}. By collecting data from each pulsar in our sample we have made histograms of their S/N, scaled to the HTRU medlat integration time and taking possible position offsets into account, and noted the average S/N from these observations as well as the catalogue S/N calculated from the catalogue flux density and eq \ref{eq:smin}. 
These histograms are shown in Fig \ref{fig:snrhist} and the corresponding values are listed in Table \ref{tab:mergetab}. In some cases it is obvious that the catalogue S/N value differs significantly from the observed values and cannot be equated to the true flux density, even if scintillation is taken into account. Some of the pulsars have a lot higher S/N values than indicated by catalogue values. See e.g. PSR\,J1125--6014, which the catalogue predicts from eq \ref{eq:smin} should have an observed S/N of only 4 but the lowest S/N of the archival data is $\sim$ 11 and the average number is 45. Similar statements can be made for e.g. PSR\,J1216--6410, PSR\,J1751--2857 and PSR\,J1843--1113. There are also cases where the opposite is true, and the catalogue flux predicted S/N is a lot larger than the average and even top values of the archival observations, e.g. for PSR\,J0900--3144 and PSR\,J1732--5049. 

To calculate luminosities of the sample MSPs, we have chosen to use the average values of the S/N from the archival observations, in combination with the periods and widths of the pulsars to calculate flux densities. In turn, these flux values were then used together with the DM derived distances\footnote{Calculated using the NE2001 model \citep{cor02}} to the pulsars to calculate their radio luminosities at an observing frequency band centred at 1.4 GHz. The values for this calculation are listed in Table \ref{tab:mergetab}.
For some of the pulsars only a few archival observations were found, not enough to get a statistically significant average value. However, in order to be consistent in our analysis we have still used our derived average values rather than the catalogue values.

\begin{figure}
	\begin{center}
		\includegraphics[width=8cm]{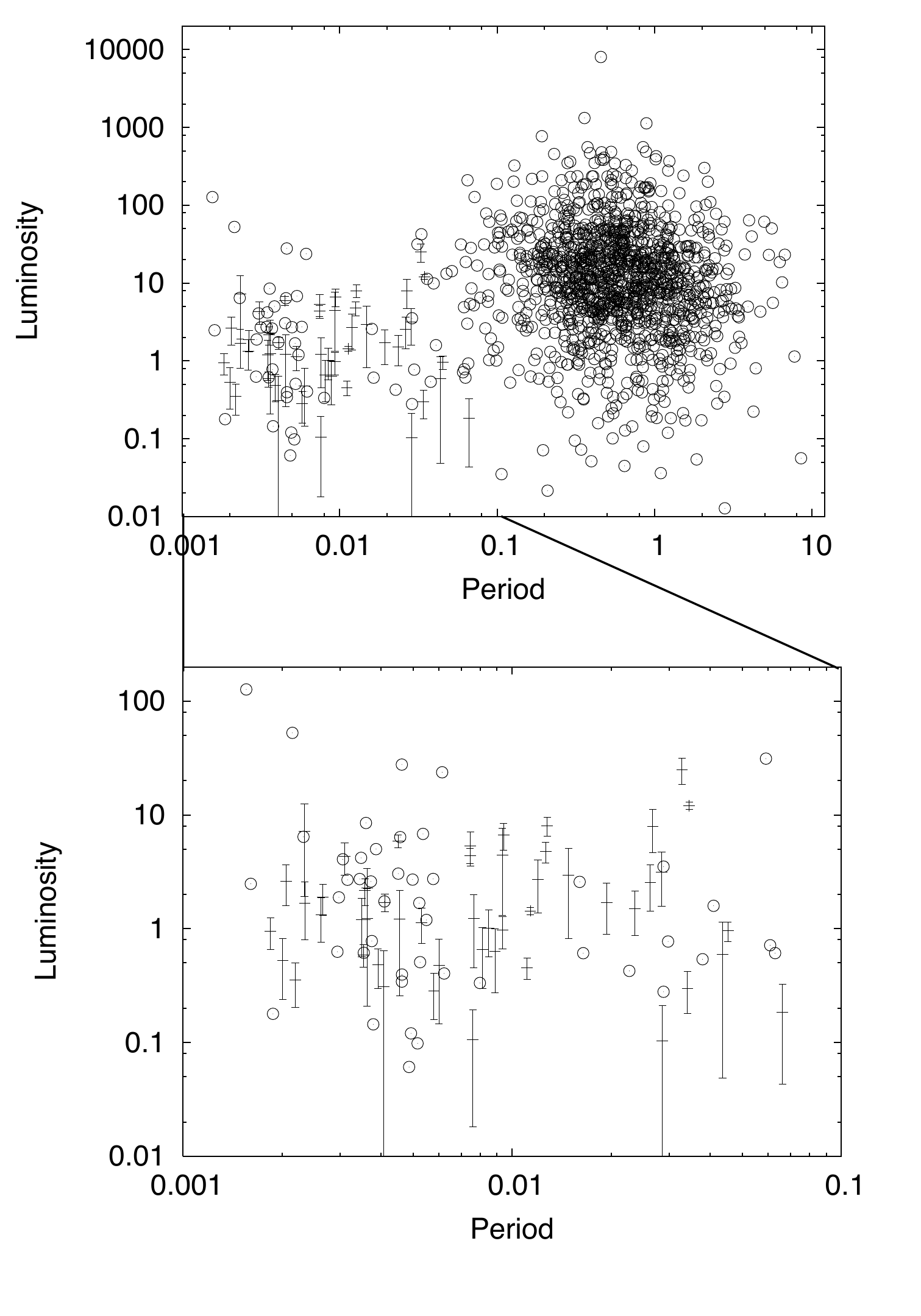}
	\caption[Luminosity - Period distribution]{The upper panel shows the luminosity - period distribution for all pulsars in the pulsar catalogue. The lower panel shows only the known recycled pulsars. In both panels, the circles represent known pulsars with values from the pulsar catalogue and the crosses represent pulsars used in this analysis with luminosity values calculated from archival data at an observing frequency of 1400\,MHz. The errors on the luminosity for the pulsars in our sample are one standard deviation with values given in Table \ref{tab:mergetab}.}
	\label{fig:lump}
	\end{center}
\end{figure}

The luminosities for the pulsars in our sample are plotted against their periods together with other known pulsars in Fig \ref{fig:lump}. The top panel includes all non-globular-cluster pulsars from the pulsar catalogue and the bottom panel shows only the recycled pulsars. The ordinary slow pulsars in the top panel ($P \gtrsim 0.1$\,s) are collected in a distinct cluster centred around $P \sim 0.5$\,s and $L \sim 50$\,mJy\,kpc$^2$. 
The sample pulsars from this analysis (represented by crosses) show no trends between luminosity and period, neither when combined with other known MSPs (represented by circles) nor by themselves.

\section{Model}
In this analysis we are using the so-called scale factor method to make an estimate of the Galactic MSP population by considering the 50 pulsars from the southern HTRU medlat survey region listed in Table \ref{tab:mergetab}. 
By ignoring pulsars below our luminosity threshold from other surveys we will only obtain a lower limit on the population above a certain luminosity threshold. 

\subsection{The Scale Factor Method}
\label{ch:scalefactor}
The scaling factor, $\xi$, is defined as the ratio of the entire weighted volume of the Galaxy to the local volume in which a pulsar is detectable. That is, the scaling factor is given by
\begin{equation}
	\xi = \frac{\int \int \int_{\rm R, z, \phi} \rho_{\rm R}(R) \rho_{\rm z}(z) R\,dR\,d\phi\,dz}{\int \int \int_{\rm R, z, \phi} \eta \rho_{\rm R}(R) \rho_{\rm z}(z) R\,dR\,d\phi\,dz}
	\label{eq:scalefactor}
\end{equation}
where $\rho_{\rm R}$ is the pulsar space density in Galactrocentric radius, $R$, and $\rho_{\rm z}$ is the density as function of the height over the Galactic plane, $z$ \citep{viv81,lor08}. The pulsar space density distribution is assumed to be uniform in Galactic azimuth position, $\phi$. The detection parameter $\eta$ depends on the simulated pulsar's coordinates as well as pulse period, $P$, and luminosity, $L$. It will return the value 1 if a pulsar with its parameters is detectable by the survey in question, and zero otherwise. 
Hence, the scaling factor is a function of pulse period and luminosity and will return a higher value for pulsars with short period and/or low luminosity, which have smaller detectable volumes. 
The model also assumes that $P, L, R$ and $z$ are independent of each other. Except for a very weak correlation between $P$ and $z$ no significant dependence exist between these quantities \citep{lor09}, and we disregard the $P-z$ correlation in this work.

Usually in this method, a separate Monte Carlo simulation is run for each known pulsar to calculate $\xi$ by evaluating the integrals in eq \ref{eq:scalefactor} using the known $P$ and $L$ values for that particular pulsar. 
In practice, for this paper, we create a large number of pulsars identical to the known pulsar and place them in the Galaxy with a reasonable distribution. After taking the inverse square law and pulse broadening into account, the number of detectable pulsars for different surveys is recorded and the scale factor is set to be the number of simulated pulsars divided by the number of detections. 

The true total number of pulsars in the Galaxy ($N_{\rm G}$) depends not only on the scale factors but also on the pulsar beaming fraction ($f$) as
\begin{equation}
	N_{\rm G} = \displaystyle\sum_{i=1}^{N_{\rm known}}{\frac{\xi_{\rm i}}{f_{\rm i}}}
\label{eq:ntot}
\end{equation}
where $N_{\rm known}$ is the number of observed pulsars used in the analysis \citep{viv81}. This gives the total number of pulsars in the Galaxy over a limiting luminosity $L_{\rm min}$, since this model is ignorant of sources below the weakest pulsar used as input. 

The beaming fraction correction in this equation originates from the finite size of the pulsar beam in the pulsar emission model and is the probability that the pulsar beam sweeps past an arbitrary observer. The beaming fraction is given by
\begin{equation}
	f = (1 - \cos{\theta}) + \left( \frac{\pi}{2} - \theta \right) \sin{\theta}
	\label{eq:bfrac}
\end{equation}
where $\theta$ is the half-angle of the emission cone \citep{emm89}.
Assuming a circular beam with a width of $\sim 10^{\circ}$ and a randomly distributed inclination angle between the spin and magnetic axes, $f$ would be approximately 20\% \citep{tay77}. However, it is known from observations that pulsars with shorter periods in general have larger beams and hence larger beaming fractions than slower pulsars \citep[see e.g.][]{nar83,tau98}. For millisecond pulsars the beaming fraction is believed to be between $0.4<f<1.0$ \citep{kra98,cam00} and likely in the upper range, closer to unity \citep{hei05}. Throughout this analysis we will assume a beaming fraction of 1. This is reasonable since the limited number of input pulsars only provides a lower limit on the total population of MSPs. 
The definition of the luminosity used here implies a beaming fraction of 0.08, and hence the luminosity is being rescaled to reflect our assumption of a beaming fraction $f=1$. 

For small samples, the detected pulsars are likely to be biased towards brighter sources, and in that case this analysis will underestimate the true underlying population. However, as long as $N_{\rm known}$ is large enough ($N_{\rm known} \gtrsim 10$), the scale factor model has proven to give reliable predictions \citep{lor93}.

\subsection{Analysis}
We have used the program {\it psrevolve}\footnote{Developed by F. Donea and M. Bailes, originally based on work by D. Lorimer. http://astronomy.swin.edu.au/$\sim$fdonea/psrevolve.html} to generate pulsars randomly spread out in the Galaxy and check if they are detectable in a number of radio pulsar surveys. 
To start with, we simulated 150,000 pulsars with periods, pulse widths and luminosities identical to each of the input pulsars in Table \ref{tab:mergetab}. These were spread out in the Galaxy assuming a Gaussian radial distribution with radial scale length $R$ = 4.5 kpc and z-values from a Gaussian distribution with a root mean square height $z$ = 500\,pc. The scale height was later varied to investigate the z-height of the underlying population. 

A snapshot approach was used, in which it is checked whether a pulsar is detectable by different pulsar surveys directly where it is placed by the simulation. For each of the surveys used in this analysis, we have created a database of the coordinates for each observation. We then define the survey region for each survey to be the area of the sky that is covered by at least one of the survey beams. This makes the analysis more accurate than simply defining the survey region as the approximate areas limited by $l$ and $b$ that are often given as survey parameters. It also gives us the advantage of knowing in which of the beams in the Multibeam receiver a pulsar was detected, as well as the exact offset of the pulsar's position to the centre of the beam. Since the telescope beams are not uniformly sensitive, neither in comparison to each other nor over the beam field of view (see Table \ref{tab:MBspecs} for details), we use this information when calculating the sensitivity for the different surveys.

To decide if a pulsar is detected or not, the program checks a few different conditions for each survey. The first of these conditions is the location of the pulsar. If the pulsar's position on the sky is outside of the survey region, the pulsar is marked as undetected. 

If the pulsar instead is inside the survey region we continue to check the second condition: the broadening of the pulse width due to propagation through the interstellar medium. 
This is done by calculating the effective pulse width for each pulsar and survey as a function of the pulsar's intrinsic pulse width and DM as well as the survey sampling time. The effective pulse width at the observer is given by:
\begin{equation}
	W_{\rm e} = \sqrt{W_{\rm i}^2 + \left( 1 + \frac{\rm DM^2}{\rm DM_0^2} \right) t_{\rm samp}^2 + t_{\rm scatter}^2}
\end{equation}
where $W_{\rm i}$ is the intrinsic pulse width, $t_{\rm samp}$ is the survey sampling time and $t_{\rm scatter}$ is the scattering time scale (in ms) given by:
\begin{equation}
	t_{\rm scatter} = 10^{-4.62+1.14\log(\rm DM)} + 10^{-9.22+4.46\log(\rm DM)} 
\end{equation}
first modelled by \cite{bha92} at 400\,MHz and scaled to the appropriate frequency for each survey after a $\nu^{-4.4}$ scaling law \citep{rom86}.
The DM in the model is calculated using the position of the simulated pulsar in the Galaxy and the NE2001 model \citep{cor02}. 
DM$_{\rm 0}$ is the dispersion smearing due to the finite sampling time and channel bandwidth in each survey, such that the smearing of the pulse in one channel is equal to the sampling time.
If the pulse width at the observer is larger than the pulse period, the pulsar is marked as undetected, and otherwise a third condition is checked. 

The third and last condition is that the flux density of the pulsar must be larger than the lower flux density limit of the survey, $S_{\rm min}$, as calculated by eq \ref{eq:smin} assuming the minimum detectable signal to noise S/N$_{\rm min}$ = 10. In this calculation also the gain of the observing beam and the position offset of the pulsar to the centre of the beam are taken into account.
If a simulated pulsar has passed all of these tests, it is marked as detected in that particular survey. 

When all of the 150,000 simulated pulsars have gone through the detection tests, we check how many of the pulsars would be detected by the HTRU medlat survey. To estimate the scale factor for each pulsar, the number of simulated pulsars is divided by the number of HTRU medlat detections. Also the detections for each of the Parkes 20-cm surveys are noted and scaled after the number of HTRU medlat detections. This entire process is run 10 times for each of 5 different z-heights: z = 100, 250, 500, 750 and 1000 pc. 

In addition, we have simulated pulsars with periods shorter than any currently known pulsar, to see if we would be able to detect these pulsars if they exist in the Galaxy. 
By creating pulsars of different periods ($P = 1.5, 1.0, 0.5$\,ms) and luminosities ($L = 0.1, 1.0, 10.0$\,mJy\,kpc$^2$), and with a pulse width of 0.10$P$, we can put limits on how many pulsars of each of these different kinds could exist in our Galaxy before we would detect one. 
In this analysis we started with simulating 150,000 of each hypothetical short period pulsar and after the simulation was finished we checked how many of these pulsars the HTRU medlat survey would have found. We then rescaled the input numbers and reran the simulation until exactly one pulsar was found. This is our upper limit on the population of pulsars with these particular properties. This process is then run 20 times to get average values on the population numbers.

\section{Results}
\input{tables/scalefactors_ver2}

The resulting scale factors for each of the sample pulsars are shown in Table \ref{tab:scalefactors}, together with the total numbers of the Galactic population of millisecond pulsars for each of the z-heights. One of the pulsars, PSR\,J1406--4656, contributes to almost half of the total population. This is the second weakest pulsar in our sample. The weakest one, PSR\,J1439--5501, has a luminosity of only $L = 0.1\,$mJy\,kpc$^2$, and as such puts the limit on the minimum luminosity we can consider in this study.
For z = 500 pc (which is a more likely value of the real z-height, see Sec \ref{sec:scaleheightdisc} for discussion) the fraction of such low luminosity pulsars to the Galactic population is about 65\%. 

From the 10 simulation runs of each scale height we can calculate average numbers of detections for each of the surveys that we are considering. These numbers are shown in Table \ref{tab:avedet}. The difference in the {\it psrcat} values and the {\it all detections} values comes from the extra search we have made of the medlat discoveries in the PM and SIL surveys, explained in Sec \ref{sec:mspdataset}. 
The simulated detections are scaled to return the right number of HTRU medlat detections. 

Average scale factors from the analysis of pulsars with periods $P \lesssim$\,1.5\,ms are shown in Table \ref{tab:smallp}. 
These values are calculated as averages over 20 runs of the simulation. Here we used a z-height of 500\,pc.

\begin{table}
  \caption[Average detection numbers]{Average numbers of detected pulsars over 10 simulation runs, scaled to the number of medlat detections. The detection values are given for two archival surveys (PM and SIL) and for the three sub-surveys of the southern HTRU survey (medlat, deep and hilat). See Sec \ref{sec:subsurveys} for specifics on the deep and hilat surveys. The ``psrcat" values are the detections listed in the pulsar catalogue and ``all detections" represents the extra search as described in Section \ref{sec:mspdataset}.}
  \begin{center}
  \begin{tabular}{l c c c c c}
	\hline 
	\hline \\[-1.5ex]
	z-height [pc] & PM & SIL & Medlat & Deep & Hilat\\
	\hline \\[-1.5ex]
	100   & 42 & 6 & 48 & 118 & 10\\
	250   & 29 & 9 & 48 & 86   & 21\\
	500   & 23 & 9 & 48 & 68   & 42\\
	750   & 22 & 9 & 48 & 62   & 56\\
	1000 & 21 & 8 & 48 & 59   & 66\\
	\hline \\[-1.5ex]
	psrcat & 14 & 12 & 48 & -- & --\\
	{\bf all detections} & {\bf 19} & {\bf 15} & {\bf 48} & {\bf --} & {\bf --}\\
	\hline
	\hline
  \end{tabular}
  \label{tab:avedet}
  \end{center}
\end{table}

\begin{table}
  \caption[Scale factors for pulsars with $P \lesssim$ 1.5\,ms]{Scale factors for simulated pulsars with $P \lesssim$ 1.5\,ms over 20 runs. A z-height of 500\,pc was used. The period, P, is given in ms and the luminosity, L, is given in mJy kpc$^2$ at an observing frequency of 1400\,MHz. The errors stated are standard deviations given in parentheses on the last quoted digit.}
  \begin{center}
  \begin{tabular}{r c c c }
	\hline
	\hline
	P  & L\,=\,0.1 & L\,=\,1.0 & L\,=\,10.0 \\ 
	\hline
	 1.5  & 28000(19000) & 800(540) & 30(20)\\
	 1.0  & 25000(17000) & 900(610) & 34(23)\\
	 0.75 & 35000(24000) & 1200(780) & 56(38)\\
	 0.5  & 38000(26000) & 1100(760) & 61(41)\\
	\hline
	\hline
  \end{tabular}
  \label{tab:smallp}
  \end{center}
\end{table}

\section{Discussion}
\label{sec:msppopdiscussion}

\subsection{S/N Variations}

Scintillation effects could result in varying values of the S/N of pulsar observations, in particular for sources with low values of DM. However, the large variations of S/N in many of the histograms in Fig \ref{fig:snrhist}, for both high- and low-DM pulsars, are surprising. Some of this variation could be due to scintillation, but it could also mean that the observing system noise fluctuates more than expected or that very thorough RFI removal is more important than assumed here.

\begin{figure}
	\begin{center}
		\includegraphics[width=8cm]{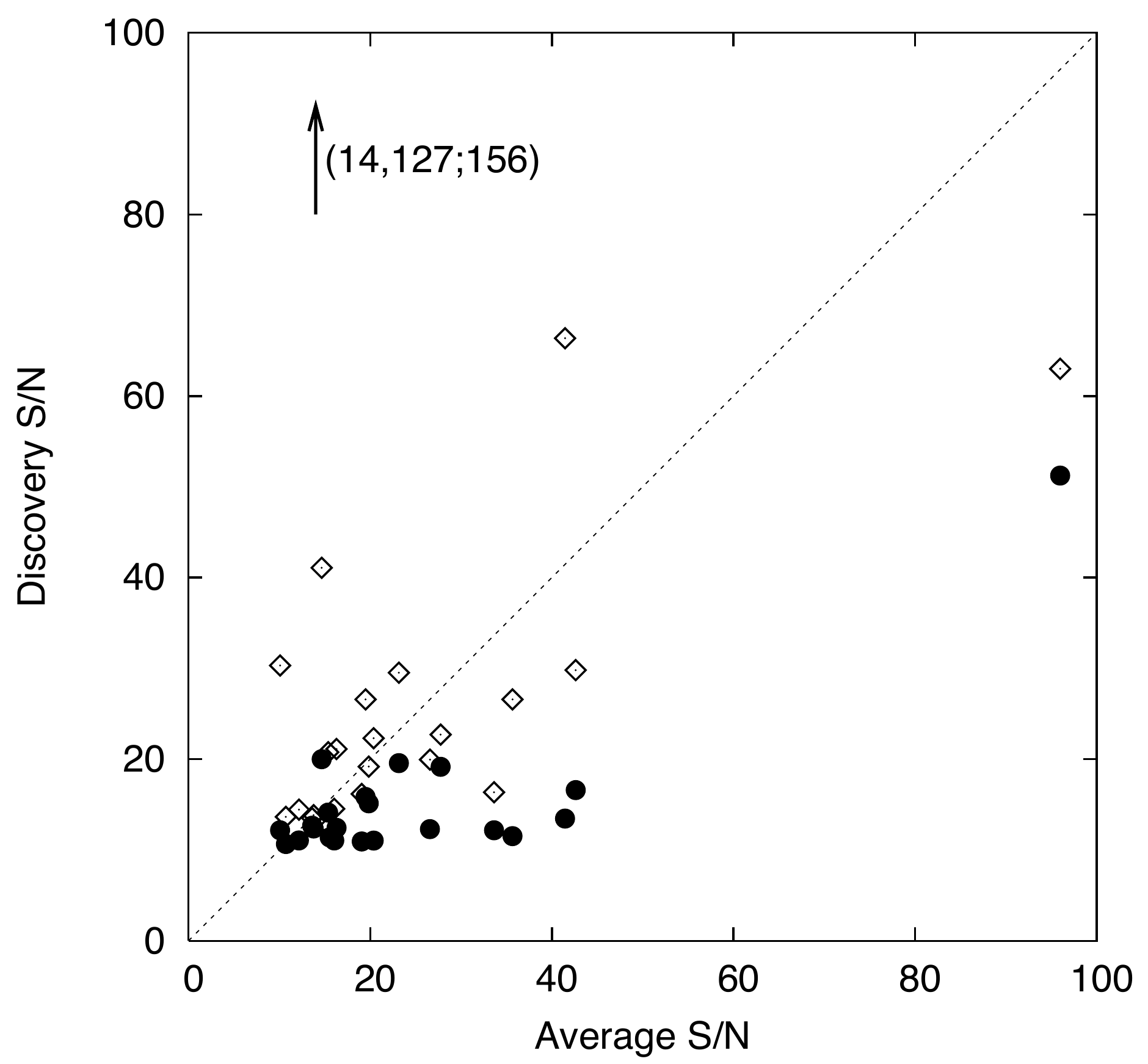}
	\caption[Comparison of the discovery S/N to the average S/N]{Comparison of the discovery S/N to the average S/N of the HTRU medlat discoveries in the sample. The black bullets represent the original S/N values reported by the processing pipeline and the open diamonds show the corresponding values after corrections for beam gain and position offset. The dashed line, y=x, is plotted for reference. The vertical arrow points to PSR\,J1406--4656, which has such a high difference in discovery S/N to average S/N that is ends up beyond the plot region.}
	\label{fig:discsnr}
	\end{center}
\end{figure}

When creating the S/N histograms, the discovery observations for the HTRU medlat pulsars were omitted. A comparison between the discovery S/N and the calculated average S/N is shown in Fig \ref{fig:discsnr}. The black bullets show the S/N of the discovery recorded by the processing pipeline. These values are in general lower than the average values, which can partly be explained by the offsets in the discovery position to the true position of the pulsar. In addition, the gain of the beams in the multibeam receiver varies between beams (as shown in Table \ref{tab:MBspecs}) and would affect the discovery S/N of the pulsars more than the average S/N since timing observations usually are performed with the centre beam. 
The open diamonds in Fig \ref{fig:discsnr} show the corresponding S/N after these effects have been corrected for. 
The dashed line represents the equality of the two S/N values, and the arrow points to a pulsar (PSR\,J1406--4656) outside of the plot with a much higher discovery S/N than average S/N. The spread of the diamonds around the equality line might be explained by scintillation effects. 

The large variations in S/N of some sources raises the question of how many pulsars exist in the Galaxy that we would detect if they were all scintillating up at the time of the observation. The modelling used in this analysis does not take scintillation into account, and this result highlights the importance of including scintillation effects in future MSP population modelling.

\subsection{Choice of Scale Height}
\label{sec:scaleheightdisc}

\begin{figure}
	\begin{center}
		\includegraphics[width=8cm]{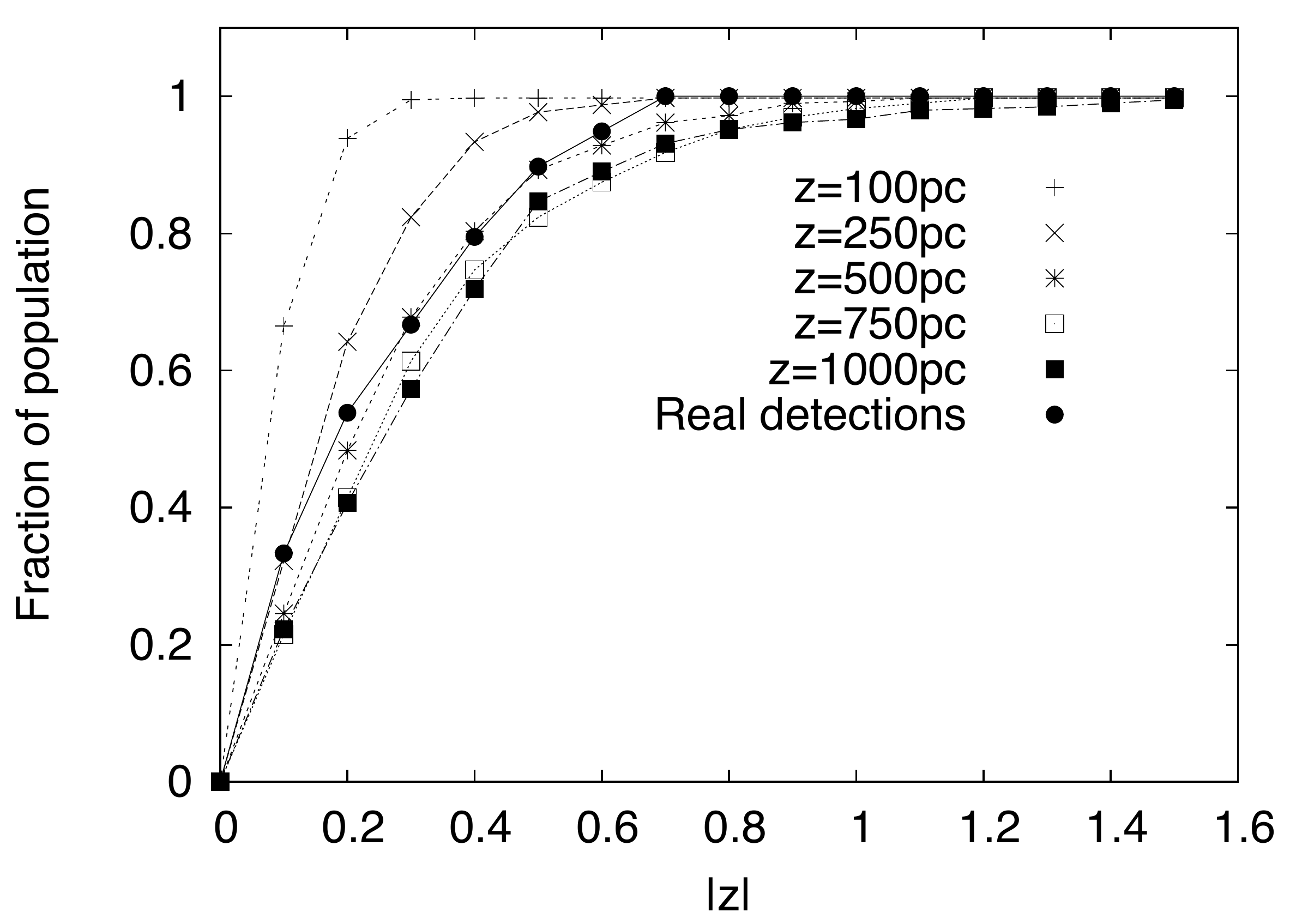}
	\caption[Cumulative histogram of scale heights]{Cumulative histogram of z-heights of simulated medlat detections together with the real medlat detections. Our best-fitting simulation was that with a scale height of 500\,pc.}
	\label{fig:cumulhist}
	\end{center}
\end{figure}

In general, the scale height of the observed MSP population is larger than the scale height of the slow pulsars, which are observed to be more tightly clustered around the Galactic plane. This is because the slow pulsars decay in luminosity rapidly with age before they manage to move to large z-heights. 
In addition, it is harder to find low-luminosity pulsars at higher z-heights, because of the large distances to those pulsars, and hence the known sample is biased towards MSPs at lower z-heights. 
This implies that the underlying Galactic population of MSPs have a larger scale height than the currently known sample, and therefore we cannot use the known z-heights of the pulsars directly in the simulation. 
Instead we have chosen to give each simulated pulsar a z-value from a Gaussian distribution with a root mean square height, $z$, and run the simulation for 5 different values of $z$ (100, 250, 500, 750 and 1000 pc). 
We can compare the z-heights for the simulated pulsar detections from these runs to the z-heights of the input pulsars to determine the most likely underlying scale height. This analysis is however only as accurate as the assumptions made on the other parameters of the simulation. 
In an attempt to avoid the bias of small number statistics at low luminosities, we have chosen to only include pulsars with $L >$\,0.5\,mJy kpc$^2$ in this analysis. 
Ten simulated survey detections of each simulated pulsar were collected and used to create cumulative histograms for each input z-height parameter. These are plotted in Fig \ref{fig:cumulhist} together
with the z-values for the real medlat detections. From this plot it seems like $z$\,=\,500\,pc best describes the real population. This result is strengthened by a Kolmogorov-Smirnov (K-S) test, which calculates the probability that the real and simulated pulsars are drawn from the same distribution. In the case of $z$\,=\,500\,pc the K-S test returns a probability of $\sim$42\%, while for $z$\,=\,250\,pc and $z$\,=\,750\,pc it returns $\sim$18\% and $\sim$17\% respectively. Hence we have chosen to use a z-height of $z$\,=\,500\,pc for the remaining part of this discussion.

\subsection{Luminosity Distribution}
\begin{figure}
	\begin{center}
		\includegraphics[width=8cm]{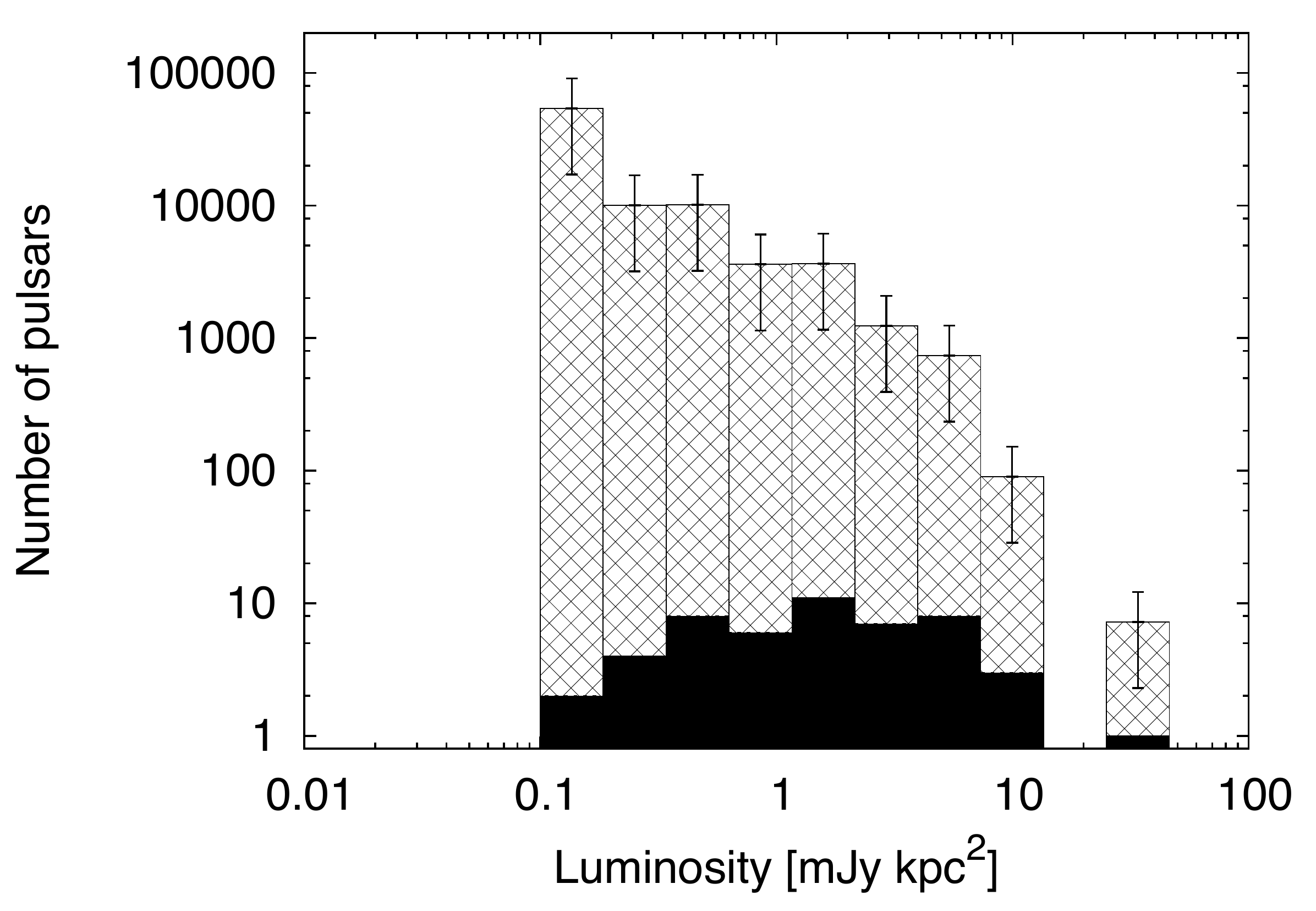}
	\caption[Luminosity histogram]{Luminosity histogram for the simulated Galactic population of MSPs over a limiting luminosity of $L$\,=\,0.1\,mJy\,kpc$^{2}$ at 1400\,MHz, assuming a z-height of $z=$\,500\,pc. The checked bars in this histogram show the simulated population and the filled bars show the numbers of real pulsars in each bin. The error bars show one standard deviation.}
	\label{fig:lumhist}
	\end{center}
\end{figure}

By using the scale factors in Table \ref{tab:scalefactors} and the luminosity for each of the pulsars, we can create a luminosity histogram for the simulated population. This can give us an idea of the underlying luminosity distribution. In Fig. \ref{fig:lumhist} we have plotted these numbers assuming a z-height of $z =$\,500\,pc. The checked bars in this histogram show the simulated population and the filled bars show the numbers of real pulsars in each bin. 

A fit of a straight line to the logarithm of the luminosity distribution of the simulated population results in a slope ($d\log N/d\log L$) of --1.45$\pm0.14$ and gives a luminosity dependence
\begin{equation}
	N = a L^b \approx 3300\pm800 \left( \frac{L}{\rm mJy\,kpc^2} \right)^{-1.45\pm0.14}
\end{equation}
for our data, with standard errors given in parentheses. Similar values calculated for slow pulsars in a snapshot model results in a slope of $\sim -1$ \citep{lor93}, and hence the MSP luminosity distribution seem to possess a steeper slope than the slow pulsars. 
Since our simulation is based directly on the luminosities of the known pulsars and we have not made any assumptions on the luminosity distribution of the MSPs in the Galaxy, this analysis can not tell us anything about the luminosity distribution below the luminosity value for the weakest pulsar in our sample.

\subsection{DM Distance Errors}
\label{sec:dmdisterr}
The validity of the scale factors calculated in this analysis is highly dependent on the estimated distance to the real and simulated pulsars. In this entire analysis we have used the NE2001 model \citep{cor02} to estimate distances/DMs for the pulsars from their DMs/distances and Galactic coordinates. 

Since the luminosity of a pulsar is proportional to the square of the distance (as $L = S d^2$) a variation in distance of 30\% translates to a variation in luminosity of 60\%. We have analysed which effect the distance has on the scale factors by assuming distances to the pulsars that are 30\% higher and 30\% lower than the DM derived value and run the simulation using the new corresponding luminosities. This resulted in an average change in the Galactic population of a factor of 3, similar to what has been shown before by \cite{lor95msp}. This verifies the importance of an accurate distance model. 

The NE2001 model is a model of the structure of the ionised gas in the Galaxy \citep{cor02}. It consists of a thin disk of scale height 140\,pc associated with low-latitude HII-regions and a thicker layer of a warm ionised medium up to a scale height of 950\,pc, as well as large-scale structure of spiral arms and other known features in the interstellar medium. This model is based partly on DMs and distances to known pulsars, and can be improved by the addition of new measurements of DMs and independent distance estimates. It has been shown that the NE2001 model sometimes is unsuccessful in predicting DMs, in particular at higher galactic latitudes, a fact which is also acknowledged by  \cite{cor02}. This has also been pointed out in other studies \citep[e.g.][]{kra03}
and has led to suggestions and attempts to improve the NE2001 model \citep[see e.g.][] {ber06,sun08,gae08,sch12}. A future implementation of an improved distance model, e.g. by using the modification proposed to the scale height by \cite{gae08}, into our simulations might result in more accurate values for the Galactic population of pulsars.

\subsection{Pulsars with $P \lesssim$ 1.5\,ms}
The fastest spinning currently known pulsar is PSR J1748--2446ad with a period $P =  1.396$\,ms \citep{hes06}, located in the globular cluster Terzan 5. Outside of globular clusters the shortest period pulsar is PSR J1939+2134 with a period $P = 1.558$\,ms \citep{bac82}. This was the first MSP ever discovered and it held the record as the fastest spinning pulsar from its discovery in 1982 until 2006 when PSR J1748--2446ad was found.
The maximum spin frequency of a pulsar is determined by the equation of state (the relationship between the density and the pressure) of the internal structure of the neutron star. By requiring that the neutron star is stable against break-up due to centrifugal forces, we can get an upper limit on the spin frequency of a neutron star with a given mass, $M$, and radius, $R$, by
\begin{equation}
	M = \frac{4\pi^2\nu_{\rm k}^2}{G}R^3
\end{equation}
where $G$ is the gravitational constant and $\nu_{\rm k}$ is the Kepler or mass-shedding frequency \citep{lor05,fri89}.  The latter, $\nu_{\rm k}$, is the highest possible frequency for a star before it starts to shed mass at the equator, and hence the maximum spin frequency of a pulsar. 
This gives a theoretical limit of the maximum spin frequency of $\nu = $\,2170\,Hz for a typical neutron star of mass $M = 1.4\,{\rm M_{\sun}}$ and radius $R=10$\,km. Recent work discussing different neutron star equations of state and rapidly rotating pulsars can be found in \cite{kras08} and \cite{hae08}.

The fastest spinning pulsar in our sample is PSR J1843--1143, $P=1.85$\,ms. In an attempt to estimate how many faster pulsars could exist in the Galaxy we have simulated pulsars with four different periods and three different luminosities. The scale factors for these twelve hypothetical pulsars can be found in Table \ref{tab:smallp}. From this analysis it is clear that at a luminosity of $L=10$\,mJy\,kpc$^2$, only a very small population ($<$181 sources) of short period pulsars could exist in the Galaxy before we would detect one, even for periods as small as $P = 0.5$\,ms.

\subsection{Estimation of Detections for the Other HTRU Sub-surveys}
\label{sec:subsurveys}
In addition to the archival Parkes 20-cm surveys and the completed HTRU medlat survey, we have included the positions of finished and proposed future observations of the HTRU hilat and HTRU deep surveys in the simulation. This provides us with estimates of how many MSPs we should anticipate detecting in these surveys. 

The deep survey covers part of the same region of the sky as the medlat survey, limited by $|b| < 3.5^{\circ}$ and $-80^{\circ} < l < 30^{\circ}$, but with an eight times longer integration time of 4300\,s. At our preferred z-height of 500\,pc we estimate that this survey will detect $\sim$ 68 MSPs, including $\sim$ 51 new discoveries.

At the same z-height we expect the hilat survey to detect 42 MSPs, including $\sim$ 27 discoveries. The hilat survey covers all the sky south of a declination of $+10^{\circ}$, not included in the medlat survey (which has limits $|b| < 15^{\circ}$, $-120^{\circ} < l < 30^{\circ}$). Each pointing in the hilat survey is integrated for 270\,s, which is half the integration time of the medlat survey observations. 

In the first survey paper of the HTRU survey, \cite{kei10} simulated the estimated detection and discovery numbers for the each of sub-surveys, using the pulsar population model described by \cite{lor06} and the {\sc psrpop}\footnote{http://psrpop.sourceforge.net} software. The main difference between that simulation and the analysis described here is that \citeauthor{kei10} assumed a luminosity distribution as a power law with index --0.59 with minimum luminosity $L_{\rm min}$ = 0.1 mJy kpc$^2$. 
The \citeauthor{kei10} simulation returned a similar number of medlat detections with 48 detections (and 28 discoveries) compared to the actual numbers of MSPs detected in the survey (51 detections, 26 discoveries). For the other two sub-surveys the corresponding values of MSPs from \citeauthor{kei10} were 51 detections, 33 discoveries for the deep survey and 65 detections, 13 discoveries for the hilat survey. 
Hence, the simulation presented here predicts a similar number of MSP detections and a larger number of  discoveries for the combined HTRU survey. 

When all the HTRU sub-surveys are finished, using the numbers of new discoveries from the simulation, we estimate the total known sample of MSPs in the southern sky to 186 sources, which corresponds to 229 MSPs in the Galaxy in total. These numbers include the MSPs recently discovered by the Fermi collaboration, which at the time of writing is 43 MSPs \citep{ray12}, but not MSPs in globular clusters. 
(It is worth noting that the numbers of discoveries predicted for the hilat and deep HTRU surveys have not taken the newly discovered Fermi pulsars into account, and hence the actual numbers of discoveries might be lower. The predicted number of detected pulsars, however, should be the same.) 
At that stage, more accurate population studies of the MSPs in the Galaxy can be performed. With such a large number of MSPs known, we would gain a better understanding of the MSP luminosity function and might be able to perform a full population synthesis study with dynamical modeling of the MSPs in the Galaxy.

\section{Conclusions}
We have performed a first-order pulsar population study, by using a snapshot model to simulate the distribution of MSPs in the Galaxy.
Since many of the published luminosity values for the MSPs in our sample were found to differ from the values calculated from the S/N measured in archival observations, we calculated average luminosities for these pulsars to use as input parameters in the simulation. 

The average luminosities for the sample pulsars show no correlation with spin period, neither by themselves nor when combined with the remaining known MSP population.  
The S/N values for many of the pulsars differ greatly between the average value and that at the time of discovery. We have chosen to use the average values in this work, but since pulsar surveys are flux limited it is arguable if the discovery S/N should be an input parameter to the model as well as the average S/N. To attempt to account for this discrepancy, we stress the importance of including scintillation effects in future pulsar population studies. 

The scale factors for the 50 sample pulsars have been simulated using 5 different values of the scale height ($z$ = 100, 250, 500, 750 and 1000 pc). A K-S test between real and simulated detections results in a best fitted scale height of $z$ = 500\,pc. By omitting beaming fractions, we find a total population of $8.3 (\pm 4.2)\times 10^4$\,MSPs in the Galaxy at $z$ = 500\,pc, down to a limiting luminosity of $L_{\rm min}$\,=\,0.1\,mJy\,kpc$^{2}$ and a luminosity distribution with a steep slope of $d\log N/d\log L$ = --1.45$\pm0.14$. 
However, at the low end of the luminosity distribution, the uncertainties introduced by small number statistics are large. By omitting very low luminosity pulsars, we find a Galactic population above $L_{\rm min}$\,=\,0.2\,mJy\,kpc$^{2}$ of only $3.0 (\pm 0.7)\times 10^4$\,MSPs. This value corresponds to a birthrate of $\sim$2.5$\times 10^{-6}$ per year, which is $\sim$10 times that of the LMXBs \citep{hur10} and implies that the birthrate problem is still present. 
These population numbers refer to MSPs with periods larger than $P=1.85$\,ms, which is the shortest spin period of the pulsars in our sample.
Using the same z-height, $z$\,=\,500\,pc, we estimate the maximum number of sub-MSPs in the Galaxy to be $7.3 (\pm 5.0)\times 10^4$ at $L$\,=\,0.1\,mJy\,kpc$^{2}$.

Finally, we predict that the HTRU deep and hilat surveys will detect 68 and 42 MSPs respectively, including a total of 78 new MSP discoveries. With the large number of detected MSPs predicted at the time of completion of the entire HTRU survey, it will be possible to perform a complete population synthesis study and extend it with dynamical modelling of the MSPs in the Galaxy.

\section*{Acknowledgements}
The work presented in this paper was conducted as part of L. Levin's PhD thesis at Swinburne University and was originally written as a thesis chapter. The version presented here has been slightly modified. 
This work is supported by the ARC Centre of Excellence for All-sky Astrophysics (CAASTRO). 
The Parkes Observatory is part of the Australia Telescope, which is funded by the Commonwealth of Australia for operation as a National Facility managed by CSIRO. 
This paper includes archived data obtained through the Australia Telescope Online Archive and the CSIRO Data Access Portal (http://data.csiro.au).
The authors would like to thank the referee for helpful comments and suggestions.

\bibliography{thesis}{}

\end{document}

%% file: tables/inputvalues_ver2.tex
\begin{table*}
  \caption[Millisecond pulsar sample]{Millisecond pulsar parameters used as input numbers in the scale factor simulation, with values used to calculate the luminosity for each MSP. The pulse width is given at 50\% of the pulse amplitude in parts of pulse period. The distance is calculated from the DM with the NE2001 model \citep{cor02}. The catalogue signal to noise (S/N$_{\rm cat}$) is calculated from eq \ref{eq:smin} and values published in the pulsar catalogue, if values exist. The average signal to noise (S/N$_{\rm ave}$) is calculated from archival Parkes observation (see text for details). The errors stated are standard deviations given in parentheses on the last quoted digit. The surveys used are the PM (p), the SIL (s) and the HTRU medlat (m) surveys.}
  \begin{center}
  \begin{tabular}{l r r r r r r r r r l}
	\hline 
	\hline \\[-1.5ex]
	Pulsar & Period   & DM & Width & S/N$_{\rm cat}$ & S/N$_{\rm ave}$ & S$_{\rm 1400}$ & Distance &  L$_{\rm 1400}$  & Survey & Reference\\
                      & [ms]       &  [pc cm$^{-3}$] &  &  &  &  [mJy] & [kpc] &  [mJy\,kpc$^2$]  & & \\
	\hline \\[-1.5ex]
J0900--3144 & 11.1 & 75.7 & 0.08(2) & 258 & 106(22) & 1.6(3) & 0.54 & 0.5(1) & m & \cite{bur06hilat}\\
J1017--7156 & 2.3 & 94.2 & 0.028(2) & 106 & 96(70) & 0.8(6) & 3.0 & 7.2(53) & m & \cite{kei12}\\
J1045--4509 & 7.5 & 58.1 & 0.11(1) & 124 & 78(26) & 1.4(5) & 2.0 & 5.3(18) & s,m & \cite{bai94}\\
J1056--7117 & 26.3 & 92.8 & 0.33(3) & -- & 11(5) & 0.4(2) & 2.6 & 2.5(11) & m & \cite{ng13}\\
J1125--5825 & 3.1 & 124.8 & 0.10(6) & 25 & 36(11) & 0.6(2) & 2.6 & 4.3(14) & m & \cite{bat11htru}\\
J1125--6014 & 2.6 & 53.0 & 0.054(4) & 4 & 45(19) & 0.6(3) & 1.5 & 1.3(6) & p,m & \cite{fau04}\\
J1157--5112 & 43.6 & 39.7 & 0.05(4) & -- & 31(28) & 0.4(3) & 1.3 & 0.6(5) & s & \cite{edw01relbin}\\
J1216--6410 & 3.5 & 47.4 & 0.056(5) & 4 & 25(6) & 0.33(8) & 1.3 & 0.6(1) & p,m & \cite{fau04}\\
J1227--6208 & 34.5 & 363.0 & 0.023(2) & -- & 20(1) & 0.18(1) & 8.3 & 12.1(9) & p,m & \cite{tho12}\\
J1337--6423 & 9.4 & 260.3 & 0.07(1) & 20 & 16(4) & 0.26(7) & 5.1 & 6.7(17) & m & \cite{kei12}\\
&&&&&&&&&& \\  
J1405--4656 & 7.6 & 13.9 & 0.15(1) & -- & 14(12) & 0.3(3) & 0.58 & 0.11(9) & m & \cite{tho12}\\
J1420--5625 & 34.1 & 64.6 & 0.033(4) & 13 & 13(5) & 0.13(5) & 1.5 & 0.3(1) & p,m & \cite{hob04}\\
J1431--4715 & 2.0 & 59.4 & 0.082(7) & -- & 14(8) & 0.2(1) & 1.6 & 0.5(3) & m & \cite{tho12}\\
J1431--5740 & 4.1 & 131.2 & 0.06(1) & -- & 19(3) & 0.26(5) & 2.5 & 1.7(3) & m & \cite{bur12}\\
J1435--6100 & 9.3 & 113.7 & 0.023(3) & 27 & 22(7) & 0.21(7) & 2.2 & 1.0(3) & p,m & \cite{cam01}\\
J1439--5501 & 28.6 & 14.6 & 0.05(2) & 33 & 22(23) & 0.3(3) & 0.60 & 0.1(1) & p,m & \cite{fau04}\\
J1446--4701 & 2.2 & 55.8 & 0.036(7) & 36 & 16(7) & 0.17(7) & 1.5 & 0.4(2) & m & \cite{kei12}\\
J1454--5846 & 45.2 & 116.0 & 0.06(1) & 15 & 13(2) & 0.20(4) & 2.2 & 1.0(2) & p & \cite{cam01}\\
J1502--6752 & 26.7 & 151.8 & 0.10(2) & 39 & 27(11) & 0.5(2) & 4.2 & 7.9(33) & m & \cite{kei12}\\
J1525--5544 & 11.4 & 126.8 & 0.042(3) & -- & 20(2) & 0.26(2) & 2.4 & 1.4(1) & p,m & \cite{ng13}\\
&&&&&&&&&& \\  
J1529--3828 & 8.5 & 73.6 & 0.13(2) & -- & 10(4) & 0.21(9) & 2.2 & 1.0(5) & m & \cite{ng13}\\
J1543--5149 & 2.1 & 50.9 & 0.11(2) & 31 & 19(8) & 0.4(2) & 2.4 & 2.6(10) & m & \cite{kei12}\\
J1545--4550 & 3.6 & 68.4 & 0.037(5) & -- & 43(11) & 0.5(1) & 2.1 & 2.2(6) & s,m & \cite{bur12}\\
J1603--7202 & 14.8 & 38.0 & 0.084(1) & 274 & 141(102) & 2.2(16) & 1.2 & 3.0(21) & s,m & \cite{lor96}\\
J1618--39 & 12.0 & 117.5 & 0.16(2) & -- & 15(7) & 0.4(2) & 2.7 & 2.7(13) & s,m & \cite{edw01rec}\\
J1622--6617 & 23.6 & 87.9 & 0.028(3) & 59 & 34(14) & 0.3(1) & 2.2 & 1.5(6) & s,m & \cite{kei12}\\
J1629--6902 & 6.0 & 29.5 & 0.06(1) & -- & 40(28) & 0.5(4) & 0.96 & 0.5(3) & s,m & \cite{edw01rec}\\
J1708--3506 & 4.5 & 146.8 & 0.17(1) & 12 & 28(4) & 0.8(1) & 2.8 & 5.9(8) & p,m & \cite{bat11htru}\\
J1719--1438 & 5.8 & 36.8 & 0.059(8) & 15 & 15(6) & 0.19(9) & 1.2 & 0.3(1) & s,m & \cite{bai11}\\
J1721--2457 & 3.5 & 47.8 & 0.25(3) & 18 & 22(12) & 0.7(4) & 1.3 & 1.2(6) & s,m & \cite{edw01rec}\\
&&&&&&&&&& \\  
J1729--2117 & 66.3 & 35.0 & 0.03(1) & -- & 15(12) & 0.2(1) & 1.1 & 0.2(1) & m & \cite{tho12}\\
J1730--2304 & 8.1 & 9.6 & 0.123(2) & 183 & 111(61) & 2.4(13) & 0.53 & 0.7(4) & p,s,m & \cite{lor95msp}\\
J1731--1845 & 2.3 & 106.5 & 0.04(1) & 54 & 23(12) & 0.3(1) & 2.6 & 1.7(9) & m & \cite{bat11htru}\\
J1732--5049 & 5.3 & 56.8 & 0.057(3) & 97 & 42(15) & 0.6(2) & 1.4 & 1.1(4) & s,m & \cite{edw01rec}\\
J1744--1134 & 4.1 & 3.1 & 0.035(2) & 317 & 172(184) & 1.8(19) & 0.42 & 0.3(3) & s,m & \cite{bai97}\\
J1745--0952 & 19.4 & 64.5 & 0.08(1) & 24 & 32(15) & 0.5(2) & 1.8 & 1.7(8) & s,m & \cite{edw01rec}\\
J1751--2857 & 3.9 & 42.8 & 0.039(4) & 3 & 21(8) & 0.4(2) & 1.1 & 0.5(2) & p,m & \cite{sta05}\\
J1755--3715 & 12.8 & 167.4 & 0.31(3) & -- & 14(3) & 0.5(1) & 3.9 & 8.0(15) & m & \cite{ng13}\\
J1756--2251 & 28.5 & 121.2 & 0.027(2) & 45 & 39(19) & 0.5(3) & 2.5 & 3.2(16) & p,m & \cite{fau04}\\
J1757--5322 & 8.9 & 30.8 & 0.049(6) & -- & 59(33) & 0.7(4) & 0.96 & 0.6(4) & s,m & \cite{edw01relbin}\\
&&&&&&&&&& \\  
J1801--1417 & 3.6 & 57.2 & 0.076(6) & 10 & 59(28) & 1.0(5) & 1.5 & 2.3(11) & p,m & \cite{fau04}\\
J1801--3210 & 7.5 & 177.7 & 0.08(1) & 12 & 16(2) & 0.27(4) & 4.0 & 4.4(7) & p,m & \cite{bat11htru}\\
J1802--2124 & 12.6 & 149.6 & 0.020(4) & 69 & 49(10) & 0.6(1) & 2.9 & 4.8(10) & p,m & \cite{fau04}\\
J1804--2717 & 9.3 & 24.7 & 0.6(3) & 5 & 97(68) & 7.3(51) & 0.78 & 4.5(31) & p,m & \cite{lor96}\\
J1810--2005 & 32.8 & 241.0 & 0.25(3) & 31 & 36(9) & 1.6(4) & 4.0 & 25.0(65) & p,m & \cite{cam01}\\
J1811--2405 & 2.7 & 60.6 & 0.05(1) & 26 & 41(13) & 0.6(2) & 1.8 & 1.9(6) & p,m & \cite{bat11htru}\\
J1825--0319 & 4.6 & 119.5 & 0.03(1) & -- & 12(10) & 0.12(10) & 3.1 & 1.2(10) & m & \cite{bur12}\\
J1843--1113 & 1.8 & 60.0 & 0.048(5) & 7 & 25(8) & 0.3(1) & 1.7 & 1.0(3) & p,m & \cite{hob04}\\
J1911--1114 & 3.6 & 31.0 & 0.10(2) & 27 & 45(37) & 0.8(7) & 1.2 & 1.2(10) & m & \cite{lor96}\\
J1918--0642 & 7.6 & 26.6 & 0.07(3) & 40 & 55(35) & 0.8(5) & 1.2 & 1.2(8) & s,m & \cite{edw01rec}\\
	\hline
	\hline
  \end{tabular}
  \label{tab:mergetab}
  \end{center}
\end{table*}

%% file: tables/scalefactors_ver2.tex
\begin{table*}
  \caption[Scale factors]{Scale factors for different z-heights of all pulsars derived in the simulation. Periods ($P$) are given in ms, luminosities ($L$) in mJy\,kpc$^2$ and z-heights ($z$) in pc.}
  \begin{center}
  \begin{tabular}{l c c | r r r r r}
	\hline 
	\hline \\[-1.5ex]
	Pulsar & $P$ & $L$ & $z$=100 & $z$=250 & $z$=500 & $z$=750 & $z$=1000\\
	\hline \\[-1.5ex]
	J0900--3144 & 11.1 & 0.5(1) & 637 & 962 & 1524 & 2181 & 2822\\
	J1017--7156 & 2.3 & 7.2(53) & 32 & 30 & 30 & 33 & 39\\
	J1045--4509 & 7.5 & 5.3(18) & 39 & 40 & 46 & 57 & 69\\
	J1056--7117 & 26.3 & 2.5(11) & 250 & 305 & 482 & 666 & 841\\
	J1125--5825 & 3.1 & 4.3(14) & 57 & 58 & 69 & 88 & 109\\
	J1125--6014 & 2.6 & 1.3(6) & 190 & 219 & 321 & 448 & 568\\
	J1157--5112 & 43.6 & 0.6(5) & 372 & 474 & 788 & 1075 & 1471\\
	J1216--6410 & 3.5 & 0.6(1) & 427 & 585 & 990 & 1435 & 1797\\
	J1227--6208 & 34.5 & 12.1(9) & 8 & 6 & 4 & 3 & 3\\
	J1337--6423 & 9.4 & 6.7(17) & 23 & 21 & 22 & 25 & 30\\
	&&&&&&&\\

	J1405--4656 & 7.6 & 0.11(9) & 7771 & 16257 & 39035 & 59250 & 67500\\
	J1420--5625 & 34.1 & 0.3(1) & 668 & 994 & 1776 & 2529 & 3537\\
	J1431--4715 & 2.0 & 0.5(3) & 692 & 1097 & 1881 & 2864 & 3532\\
	J1431--5740 & 4.1 & 1.7(3) & 131 & 143 & 196 & 271 & 339\\
	J1435--6100 & 9.3 & 1.0(3) & 138 & 151 & 214 & 287 & 379\\
	J1439--5501 & 28.6 & 0.1(1) & 3241 & 7163 & 14930 & 23214 & 34845\\
	J1446--4701 & 2.2 & 0.4(2) & 675 & 963 & 1782 & 2434 & 3355\\
	J1454--5846 & 45.2 & 1.0(2) & 220 & 260 & 402 & 550 & 777\\
	J1502--6752 & 26.7 & 7.9(33) & 19 & 19 & 20 & 24 & 28\\
	J1525--5544 & 11.4 & 1.4(1) & 120 & 133 & 178 & 243 & 307\\
	&&&&&&&\\
	
	J1529--3828 & 8.5 & 1.0(5) & 373 & 484 & 805 & 1168 & 1438\\
	J1543--5149 & 2.1 & 2.6(10) & 119 & 129 & 180 & 241 & 297\\
	J1545--4550 & 3.6 & 2.2(6) & 87 & 92 & 115 & 149 & 191\\
	J1603--7202 & 14.8 & 3.0(21) & 67 & 70 & 89 & 117 & 145\\
	J1618--39 & 12.0 & 2.7(13) & 124 & 137 & 191 & 262 & 329\\
	J1622--6617 & 23.6 & 1.5(6) & 78 & 86 & 110 & 145 & 185\\
	J1629--6902 & 6.0 & 0.5(3) & 540 & 746 & 1308 & 1845 & 2782\\
	J1708--3506 & 4.5 & 5.9(8) & 48 & 49 & 58 & 73 & 91\\
	J1719--1438 & 5.8 & 0.3(1) & 1058 & 1618 & 2838 & 4513 & 5169\\
	J1721--2457 & 3.5 & 1.2(6) & 495 & 675 & 1163 & 1658 & 2262\\
	&&&&&&&\\
	
	J1729--2117 & 66.3 & 0.2(1) & 1006 & 1690 & 3278 & 4469 & 5521\\
	J1730--2304 & 8.1 & 0.7(4) & 575 & 796 & 1284 & 1939 & 2699\\
	J1731--1845 & 2.3 & 1.7(9) & 140 & 156 & 214 & 287 & 371\\
	J1732--5049 & 5.3 & 1.1(4) & 198 & 237 & 355 & 508 & 642\\
	J1744--1134 & 4.1 & 0.3(3) & 813 & 1140 & 2157 & 3145 & 4038\\
	J1745--0952 & 19.4 & 1.7(8) & 136 & 154 & 218 & 293 & 376\\
	J1751--2857 & 3.9 & 0.5(2) & 454 & 581 & 1071 & 1381 & 1945\\
	J1755--3715 & 12.8 & 8.0(15) & 51 & 53 & 64 & 81 & 102\\
	J1756--2251 & 28.5 & 3.2(16) & 29 & 29 & 31 & 37 & 45\\
	J1757--5322 & 8.9 & 0.6(4) & 387 & 483 & 804 & 1063 & 1442\\
	&&&&&&&\\

	J1801--1417 & 3.6 & 2.3(11) & 102 & 111 & 146 & 197 & 255\\
	J1801--3210 & 7.5 & 4.4(7) & 41 & 42 & 48 & 60 & 74\\
	J1802--2124 & 12.6 & 4.8(10) & 22 & 20 & 19 & 20 & 23\\
	J1804--2717 & 9.3 & 4.5(31) & 239 & 293 & 442 & 628 & 807\\
	J1810--2005 & 32.8 & 25.0(65) & 7 & 7 & 7 & 7 & 8\\
	J1811--2405 & 2.7 & 1.9(6) & 122 & 135 & 180 & 242 & 307\\
	J1825--0319 & 4.6 & 1.2(10) & 149 & 165 & 229 & 320 & 404\\
	J1843--1113 & 1.8 & 1.0(3) & 282 & 343 & 536 & 752 & 985\\
	J1911--1114 & 3.6 & 1.2(10) & 258 & 312 & 478 & 679 & 936\\
	J1918--0642 & 7.6 & 1.2(8) & 200 & 235 & 353 & 492 & 625\\
	\hline
	\hline \\[-1.5ex]
	TOTAL: & & & 23910   &   40948  &    83461  &   124448  &   156842\\
	\hline
	\hline
  \end{tabular}
  \label{tab:scalefactors}
  \end{center}
\end{table*}